\documentclass[singlecoloumn]{revtex4}

\usepackage{graphicx}
\begin{document}

\title{Hidden role of antiunitary operators in Fierz transformation}

\author{Igor F. Herbut}

\affiliation{Department of Physics, Simon Fraser University, Burnaby, British Columbia, Canada V5A 1S6}

\begin{abstract}

We show that whenever the symmetry group of a field theory commutes with one or more antiunitary operators $T$, which do not have to but may represent the reversal of physical time, the number of linearly independent contact two-body (quartic) terms is determined by the number of tensors that are even, or by the number of tensors that are odd, under such $T$. The choice depends on the sign of $T^2$ and on the statistics of the fields. The theorem enables one to circumvent the usual computation of the Fierz matrix in determining the independent interaction terms. Some physical examples of current interest in many-body physics are discussed.

\end{abstract}

\maketitle

\section{Introduction}

The problem of determining the form of a field theory allowed by symmetry has been central to many branches of many-body physics, be it the theory of classical phase transitions, elementary particle physics, or quantum condensed matter. A large symmetry group is often quite restrictive so that only a few low-order interaction terms  are possible and easy to guess. There are however situations when this is not the case, and the symmetry alone allows for several, sometimes even a large number of relevant interaction terms. It then becomes of central importance to eliminate the possible redundancy and determine the set of independent interactions. If these are, as is common, quartic in the fields, they are related by the Fierz transformations, \cite{fierz,itzykson} which represent an expression of completeness of a chosen basis in the given Hilbert space, and prescribe how to rewrite each allowed quartic term in terms of all others. The dimension of the kernel of the matrix produced by such an algorithm equals then the number of independent quartic terms,\cite{hjr} and one can from the matrix also find what these are. The ``Fierz matrix" is real, but not necessarily symmetric, and its calculation, while in principle straightforward, is also a tedious task. In the well-scrutinized electronic systems of monolayer and bilayer graphene, for example, even assuming spinless fermions the translational and rotational symmetry allows nine quartic terms, and thus calls for the computation of close to eighty elements of the  Fierz matrix.\cite{vafek} Including the electron spin simply doubles the size of the matrix that needs to be computed. It therefore appears to be of some value to think of a shortcut to the solution, which would enable one to determine the number of independent couplings and maybe even their identity without doing the full calculation.

Here we point out that such a shortcut is, somewhat unexpectedly, facilitated by an antiunitary operator that commutes with the symmetry group in the given representation, whenever such an operator exists. Assume that the fields belong to some irreducible representation of a given, continuous or discreet symmetry. Schur's lemma implies then that the only linear operator that commutes with the whole group in the representation is proportional to the unit operator. It does not forbid, however, antiunitary such an operator, and indeed the case of an antiunitary operator representing the time-reversal which commutes with the group of three-dimensional spatial rotations (in any representation) is a textbook example. The existence of such an operator is tantamount to (pseudo)reality of the representation, and it will, for example, always exist if the symmetry group derives from space rotations, which is often true in condensed matter physics. For finite groups its existence can be ascertained by the Frobenius-Schur indicator. If the field representation of the group is reducible, there could even be more than one such operator.

The main result of the present note is then the following: if the fields are bosonic (complex numbers), the number of independent quartic terms equals the number of tensors in the given representation which are {\it even} under the antiunitary operator $T$ that commutes with the symmetry group of the theory, when $T^2 =1$, i. e. when the representation is ``real". It equals the number of tensors which are odd, when $T^2 =-1$, and the representation is ``pseudoreal". For fermionic (Grassmann) fields it is the other way around. It then also follows that in reducible representations with more than one such antiunitary operator $T$, all those that have $T^2=1$ yield the same number of even tensors, which is in turn equal to the number of tensors odd under those antiunitary operators with $T^2=-1$.  Important examples of such redundancy which nevertheless yield unique conclusions are provided by the above-mentioned graphenelike systems.

In the rest of the paper we first define the general Fierz transformation for both fermions and bosons, and then introduce the ``type-II"  quartic terms which are the main tool in the proof. After presenting the proof of the theorem we discuss several physical examples which illustrate principles at work. Summary is provided at the end.

\section{Fierz transformation}

Assume a global symmetry group of linear transformations $G\times U_p (1)$, and the field $\Psi(x)$ as a d-dimensional column of complex or Grassmann numbers which transforms under $G$, and acquires a phase under $U_p (1)$. One can think of $x$ here as a coordinate, but it can be any such label. There will be $d^2$ linearly independent Hermitian matrices $X$  in the space of $d$-dimensional matrices. These can be grouped into irreducible representations of $G$, so that the bilinears $\Psi^\dagger X^a _i \Psi$ under the transformation $\Psi \rightarrow g \Psi$ with $g\in G$  transform as
\begin{equation}
\Psi(x) ^\dagger X^a _i \Psi(x) \rightarrow \sum_j c_{ij}  \Psi(x) ^\dagger X^a _j \Psi(x),
\end{equation}
while remaining invariant under a $U_p (1)$ transformation.
The upper index $a$ numerates different irreducible tensors under $G$, which are $d \times  d$ matrices, and the lower indices $i$ and $j$ numerate different components of the same tensor. For example, if $d=3$ and $G=SO(3)$, there are three such tensors: scalar, vector, and irreducible (traceless) second-rank tensor, which contain one, three, and five components, respectively. \cite{QM} Orthogonality between matrices may be defined as
\begin{equation}
\frac{1}{d} Tr X^a _i X^b _j  =  \delta_{ij} \delta_{ab}.
\end{equation}

We assume for simplicity that all the quartic terms invariant under $G\times U_p (1) $ can be written as linear combinations  of the local, call them ``type-I" (or ``particle-hole"),  terms in the form
\begin{equation}
 \sum_i (\Psi^\dagger (x) X^a _i \Psi(x))( \Psi^\dagger(x) X^a _i \Psi(x)) ,
\end{equation}
with one such term existing for each tensor $X^a$. This will typically be the case, but we will encounter exemptions among our examples, and discuss them as well. Since the set of all matrices $X^a _j$ forms a basis one can deduce the Fierz identities \cite{itzykson, hjr}
\begin{equation}
\sum_i ( \Psi^\dagger (x_1) X^a _i \Psi(x_2) ) (\Psi^\dagger (y_1)  X^a _i \Psi(y_2))
  = \frac{s}{d} \sum_{b,c,j,k,i}Tr (X^a _i X^b _j X^a _i  X^c_k) ( \Psi^\dagger (x_1) X^c _k \Psi(y_2)) ( \Psi^\dagger (y_1) X^b _j \Psi(x_2) ),
\end{equation}
where $s=1$ for complex (bosonic) fields, and $s=-1$ for Grassmann (fermionic) fields. Although the textbooks discuss the fermionic case almost exclusively, it is easy to see that the Fierz rearrangement formula is independent of statistics, and applies  equally well to complex fields, only with a different overall sign.

Under the assumption that all the $G\times U_p (1)$ - symmetric quartic terms are in the form of Eq. (3) above only the terms with $b=c$ and $j=k$ survive the summation on the right-hand side of Eq. (4). This implies that the type-I  terms are not independent of each other: there is an equation $F Q = 0$, where $Q$ is a column of all symmetry allowed quartic terms, and $F$ is a matrix produced by the Fierz transformations. The number of independent terms is given by the dimension of the kernel of the Fierz matrix $F$.\cite{hjr} Our task is to determine this number, and maybe even the identity of the independent terms,  without actually computing the matrix $F$.

\section{Type-II terms}

 The gist of our method is the observation that the bilinears such as
  \begin{equation}
  \Psi^\dagger (x) X^a_j (T \Psi(x)) = \Psi^\dagger (x) X^a_j U \Psi^* (x),
  \end{equation}
  and
 \begin{equation}
  [\Psi^\dagger (x) X^a_j (T \Psi(x))]^\dagger = s \Psi^T (x) U^{-1} X^a_j \Psi (x),
  \end{equation}
   where $T=U K$ is the antiunitary operator, with $U$ as its unitary part and $K$ standing for complex conjugation, {\it also} transform the same way as the irreducible tensor $X^a$ under the transformation $\Psi \rightarrow g \Psi$, if $T$ and all $g\in G$ commute in the given representation, i. e. if
   \begin{equation}
   U g^* U^{-1} =g.
   \end{equation}
There exist therefore other  $G\times U_p (1)$ - invariant local quartic terms in what one could call ``type-II" (or ``particle-particle") form
\begin{equation}
 \sum_i [ \Psi^\dagger (x) X^a _i (T \Psi(x)) ]  [\Psi^\dagger (x) X^a _i (T \Psi(x)) ]^\dagger    ,
\end{equation}
for each tensor $X^a$. From the ordering of the fields in the type-II terms it is evident, however, that the Fierz transformations in Eq. (4) relate them  not to the other type-II terms, but exclusively to the previous set of type-I terms. The type-II terms are therefore not truly new, but simply are  linear combinations  of the terms introduced in the previous section. The main point  is that some of the type-II terms are in fact identically {\it zero}. We then show that each such vanishing type-II term yields a linearly independent constraint on the type-I terms, so that the number of nonvanishing type-II terms equals in fact the number of linearly independent type-I terms. This number ultimately depends on the transformation property of the tensor under $T$ (even vs. odd)  and the statistics of the fields in a way that implies the result announced in the introduction.

\section{Proof}

To show the above consider the bilinear
\begin{equation}
\Psi^\dagger (x) M (T \Psi (x) ) =  \Psi^\dagger (x) M U \Psi^* (x),
\end{equation}
 where $M$ is a Hermitian $d$-dimensional matrix. By transposing we find that
 \begin{equation}
 \Psi^\dagger (x) M U \Psi^* (x)  = s \Psi^\dagger (x) U^T M^T  \Psi^* (x),
\end{equation}
with $s$ as the previously introduced sign for the statistics of the fields.
For any matrix $M$ for which it happens that $ M U = -s U^T M^T $ the bilinear  is therefore identically  zero.

Let us assume that $T^2 =t =\pm 1$, so that $ U U^* =  t$, and thus $U^* = t U^{-1}$. Since $U^\dagger = U^{-1}$, we have that $U^T = ( U^* )^{-1}$, so finally $U^T = t U$. Since $M$ is Hermitian and $M^T = M^*$, the condition for the vanishing of the bilinear is that $M U = - s t U M^*$, or rewritten,
 \begin{equation}
 M = - s t U M^* U^{-1}.
 \end{equation}
 We recognize the combination $U M^* U^{-1} = T M T^{-1}$ as precisely the matrix $M$ transformed under $T$. The result is that any bosonic, $s=1$, bilinear involving a matrix $M$ that is odd (even) under $T$ will vanish if $t=1$  ($t=-1$), and any fermionic, $s=-1$, bilinear with $M$ that is even (odd) will vanish if $t=1$ ($t=-1$). When $T$ commutes with the elements of $G$ all the components of the same tensor have the same transformation under $T$, and therefore some of the type-II terms are identically zero.

The number of independent linear Fierz constraints therefore cannot exceed the number of type-II tensors that yield vanishing bilinears in Eq. (5), but one could allow for the possibility that their number is smaller. The remaining step is therefore to prove that each vanishing type-II term implies a linearly independent constraint between the type-I terms. The number of independent quartic terms is then simply given by the difference between the total number of symmetry allowed type-I (or type-II) terms and the number of vanishing type-II terms.

To that purpose let us write some vanishing type-II term (Eq. (8)) and Fierz transform it first as
\begin{eqnarray}
0=  \sum_j [ \Psi^\dagger (x) X^a_j U \Psi^* (x)] [ \Psi^T (x) U^{-1} X^a_j \Psi (x) ] = \\ \nonumber
= \frac{s}{d} \sum_{b,c,j,k,i}Tr (X^a _i U X^b _j U^{-1}  X^a _i  X^c_k) ( \Psi^\dagger (x) X^c _k \Psi(x)) ( \Psi^T (x) X^b _j \Psi^* (x) ) \\ \nonumber
= \frac{1}{d} \sum_{b,c,j,k,i} Tr (X^a _i U X^b _j U^{-1}  X^a _i  X^c_k) ( \Psi^\dagger (x) X^c _k \Psi(x)) ( \Psi^\dagger (x) (X^b_j)^T  \Psi (x) ),
\end{eqnarray}
where  we have transposed the last factor in going from the second to the third line. Since $X^b _j $ is Hermitian, $(X^b _j)^T = (X^b _j )^*$.
Changing the matrix $(X^b _j )^* \rightarrow X^b _j$ in the sum the last line becomes
\begin{equation}
0=\sum_{b,c,j,k,i} Tr (X^a _i U (X^b _j)^*  U^{-1}  X^a _i  X^c _k) ( \Psi^\dagger (x) X^c _k \Psi(x)) ( \Psi^\dagger (x) X^b_j   \Psi (x) )
\end{equation}
One recognizes the matrix featuring under the trace
\begin{equation}
U (X^b _j)^*  U^{-1} = T X^b _j T^{-1} = \pm X^b _j,
\end{equation}
with the sign depending only on the tensor (i. e. on the upper index ``b") and not on its component (index ``j"). Every vanishing type-II term produces therefore one linear equation on the type-I terms such as
\begin{eqnarray}
 \sum_{b - even} \sum_{j,i} Tr (X^a _i X^b _j  X^a _i  X^b _j) ( \Psi^\dagger (x) X^b _j \Psi(x)) ( \Psi^\dagger (x) X^b _j   \Psi (x) ) = \\ \nonumber
 \sum_{b - odd}  \sum_{j,i} Tr (X^a _i X^b _j  X^a _i  X^b _j) ( \Psi^\dagger (x) X^b _j \Psi(x)) ( \Psi^\dagger (x) X^b_j   \Psi (x) ).
\end{eqnarray}
For that same tensor the type-I term could be, on the other hand, written as
\begin{equation}
\sum_i ( \Psi^\dagger (x) X^a _i \Psi(x) ) (\Psi^\dagger (x)  X^a _i \Psi(x))
  = \frac{s}{d} \sum_{b - all}\sum_ {j,i}Tr (X^a _i X^b _j X^a _i  X^b_j) ( \Psi^\dagger (x) X^b _j \Psi(x)) ( \Psi^\dagger (x) X^b _j  \Psi(x) ).
\end{equation}
The last two equations may then be combined into
\begin{equation}
\sum_i ( \Psi^\dagger (x) X^a _i \Psi(x) ) (\Psi^\dagger (x)  X^a _i \Psi(x)) =
\frac{2s}{d} \sum_{b}^{'}  \sum_{j,i}  Tr (X^a _i X^b _j X^a _i  X^b_j) ( \Psi^\dagger (x) X^b _j \Psi(x)) ( \Psi^\dagger (x) X^b _j  \Psi(x) ),
\end{equation}
where the primed sum is taken only over the tensors $X^b$ that have the  sign under the transformation $T$ {\it opposite} of that of the selected tensor $X^a$. Since there is one such equation for each tensor $X^a$ that yields a zero type-II term, and these by construction do not appear on the right-hand side of the last equation, such equations deriving from different tensors are manifestly linearly independent. The number of linearly independent Fierz constraints matches exactly the number of tensors that yield vanishing type-II terms. The number of independent type-I quartic terms is therefore the total number of different tensors minus the number of tensors that give vanishing type-II terms, as claimed.

\section{Examples}

  Let us now consider some examples in order of increasing complexity.

  1) The simplest example is probably $G= SU(2)$, and $\Psi(x)$ in the two-component representation. The operator $T = \sigma_2 K$ is then unique, $T^2=-1$, and there is one scalar ($1_2 $, unit matrix) and one vector ($\sigma_i$, $i=1,2,3$), even and odd under $T$, respectively. The number of independent quartic terms  is thus in both the bosonic (complex) and fermionic (Grassmann) case equal to one, and it may simply  be taken to be the usual $(\Psi^\dagger (x) \Psi(x) )^2$.

  2) $G= SO(3)$, with $\Psi(x)$ in the three-dimensional ($j=1$) representation. In the adjoint representation of the $SO(3)$,  $T=K$, unique, and $T^2=1$. There is  a scalar (even), vector (odd), and second-rank tensor (even) in the space of Hermitian $3 \times 3$ matrices, so if $\Psi$ is a complex field the number of independent quartic terms is two. They may be taken to be $(\Psi^\dagger (x) \Psi(x))^2$ and $|\Psi^T (x) \Psi(x)|^2$, for example.

  Assuming a Grassmann $\Psi(x)$ violates the spin-statistics theorem, nevertheless, such a situation could arise in solids if three bands cross at a point in the Brillouin zone.\cite{bradlyn} In the fermionic case we have then only one independent quartic term, say $(\Psi^\dagger (x) \Psi(x))^2$.

It is interesting to add the spin-1/2 degree of freedom, and consider the symmetry group to be $G=SU(2) \times SO(3)$, and two three-component fields $\Psi_\alpha(x)$, $\alpha=1,2$, transforming as a doublet under the $SU(2)$. Then $T^2 =-1$, and each of the above $3\times 3$ scalar, vector, and second-rank tensor may be multiplied by either $1_2$ (even) or $\sigma_i$ (odd) in the spin space. The total number of invariant terms is then six, and there are three even and three odd terms under $T$. The number of independent couplings is therefore equal to three, for both complex or Grassmann fields.\cite{rahul}

3) $G= SO(3)$, with $\Psi$ in the four-dimensional ($j=3/2$) representation. $T$ is unique with $T^2 =-1$. For bosonic $\Psi$ this violates the spin-statistic theorem, and we know of no physical realization. Nevertheless, since the space of $4 \times 4$ Hermitian matrices consists of a scalar (even), vector (odd), irreducible second-rank (even), and irreducible third-rank (odd) tensors under $SO(3)$, there are two independent couplings for both complex and Grassmann $\Psi$. They can be taken to be $(\Psi^\dagger (x) \Psi(x))^2$ and $(\Psi^\dagger (x) S_i \Psi(x))^2$, where $S_i$ are the generators of $SO(3)$, for example. \cite{boettcher}

4) Four-dimensional representation can also arise as a spinor representation of the group $G=SO(5)$. In that case there is a scalar ($1_4$), vector ($\gamma_a$, $a=1,...5$), and second-rank tensor representation ($i[\gamma_a , \gamma_b]/2$ ), where $\gamma_a$ represent the generators of the Clifford algebra $C(5,0)$, i. e. are five mutually anticommuting  $4\times 4$ Hermitian matrices that square to unity. Since three of the $\gamma$-matrices may be chosen as real, say $a=1,2,3$, and the remaining two as imaginary \cite{herbut-clifford, herbut-quaternion}, the unique antilinear operator $T$ that commutes with the ten generators $i[\gamma_a, \gamma_b]/2$ is $T= i\gamma_4 \gamma_5 K$, with $T^2 =-1$. Since the scalar and the vector are now even under $T$, whereas the second-rank tensor is odd, there will be two independent couplings for the Grassmann field, and only one for the complex field. The former can be taken to be $(\Psi^\dagger (x) \Psi(x))^2$ and $(\Psi^\dagger (x) \gamma_a \Psi(x))^2$, for example.\cite{janssen}

5) Assume again $G= SO(3)$, but $\Psi(x)$ in five-dimensional ($j=2$) representation, with a unique $T$ with $T^2 = 1$. Since there is now a scalar (even), vector (odd), second-rank  (even), third-rank (odd), and fourth-rank tensor (even) available in the $5 \times 5$ matrix space, for a complex $\Psi$ there  are three independent couplings, \cite{kawaguchi, boettcherPRL, boettcherPRB} whereas for Grassmann $\Psi$ there will be only two. Since for the complex $\Psi$ the difference
\begin{equation}
(\Psi^\dagger (x) \Psi(x))^2 - |\Psi^\dagger (x) \Psi(x)^*|^2
\end{equation}
can in analogy to the derivation of the Eq. (17) be shown to be the sum of two quartic terms that contain the only odd tensor bilinears, the linearly independent terms may be taken to be
\begin{equation}
(\Psi^\dagger (x) \Psi(x))^2,  |\Psi^\dagger (x) \Psi(x)^*|^2 , \sum_{i=1} ^3(\Psi^\dagger (x) S_i\Psi(x))^2
\end{equation}
where $S_i$ are the $j=2$ generators of $SO(3)$, in the representation in which $T=K$.

The results above in the case of even angular momentum $j$ and $\Psi$ in the $2j+1$-dimensional representation of the $SO(3)$ generalize into $j+1$ independent quartic terms for bosons, and $j$ independent quartic terms for fermions. The same of course follows from considering the addition of the angular momentum and the symmetry of the two-particle wave function.\cite{barton} For half-integer $j$ the number is the same for both statistics, when it equals $j + 1/2$.

6) Let us assume a four-component Grassmann $\Psi$ in a {\it reducible} representation of $G= SO(3) \times U(1)$. Assume further a Clifford algebra $C(5,0)$ of $4\times 4$ Hermitian matrices $(\alpha_1, \alpha_2, \alpha_3, \beta_1, \beta_2)$, each one with a square of unity, and $\alpha_i$ real and $\beta_i$  imaginary.\cite{herbut-clifford, herbut-quaternion} We will consider the three Hermitian generators of the $SO(3)$ to be $i \alpha_i \alpha_j$, with $i\neq j$, and the generator of the $U(1)$ to be $i\beta_1 \beta_2$. This would be the symmetry group of the Weyl fermion in condensed matter systems, where the $SO(3)$ is the group of rotations, and the $U(1)$ is related to translation.\cite{hjr}  At low energies the single-particle (Dirac) Hamiltonian may be taken to be $H= \alpha_i p_i + O(p^2)$. The above symmetry group is exact, however, and it holds beyond the leading term in momentum expansion.

We can now discern the following six groups of the sixteen Hermitean operators in the $4\times 4$ space as being irreducible representations of the above group: 1) $1_4$, which is scalar under both $SO(3)$ and $U(1)$ (``scalar-scalar", respectively), 2) $i\beta_1 \beta_2$ , (scalar-scalar), 3) $\alpha_i$, $i=1,2,3$ (vector-scalar), 4) $i\alpha_i \alpha_j$, $i\neq j$ (vector-scalar), 5) $\beta_i$, $i=1,2$ (scalar-vector), and 6) $i\beta_i\alpha_j$, $i=1,2$, $j=1,2,3$, (vector-vector).

The antiunitary operator that commutes with the symmetry group in this reducible representation is now not unique: there is
$T_1=i\beta_1 \beta_2 K$, with $T_1 ^2 =-1$, but also  $T_2 = K$, with $T_2 ^2 =1 $.
The first one happens to commute with the Hamiltonian $H$ and may be taken to represent the physical time reversal, but this is not essential for our present purpose. We want to check the assertion that even in this situation our theorem leads to the  unique answer for the number of independent quartic terms. Take first $T_1$: the above six groups of operators are even, odd, even, odd, even, odd, under $T_1$, respectively. One therefore has three independent terms, irrespectively of the assumed statistics for the field.  The same conclusion follows from considering $T_2$: the operators are then even, odd, even, odd, odd, even, respectively. In spite of the reversal of the transformation property of the last two terms the total number of even (or odd) terms remained the same.

This example also provides an exception from the requirement that all symmetric terms are in the form as in Eq. (3). There are now two different tensors with the identical transformation property under $SO(3)\times U(1)$, namely the two vector-scalars on the above list. The symmetry group $SO(3)\times U(1)$ alone therefore also allows a mixed term between them. Only if we add the timereversal and consider
$T_1 \times SO(3)\times U(1)$ as the symmetry group, the mixed term becomes prohibited. Alternatively, if one assumes the Weyl Hamiltonian to be symmetric under parity, the matrices $\alpha_i$ would be parity-odd and $i\alpha_i\alpha_j$ parity-even, which would also eliminate the mixed term. Since the Weyl Hamiltonian in a solid always respects either parity or time reversal, the mixed term is in fact forbidden.

7) The final example is the two-dimensional version of the previous one: assume Grassmann four-component $\Psi$, and the symmetry group to be only $G= U(1)\times U(1)$, with the left $U(1)$ being generated by $i\alpha_1 \alpha_2$, and the right one with $i \beta_1 \beta_2$. The left $U(1)$ is the  group of rotations in the plane, and the right one is still related to  the translations. As a physical example one may take the low-energy Hamiltonian for single-layer (spinless) graphene, $H_1=\alpha_1 p_1 + \alpha_2 p_2 +O(p^2)$,\cite{hjr} or for bilayer (spinless) graphene, $H_2 = \alpha_1 (p_1 ^2 - p_2 ^2) + 2 \alpha_2  p_1 p_2 +O(p^3)$.\cite{vafek} Both of these have in fact symmetry groups larger than $U(1) \times U(1)$, but the inclusion of the higher-order terms would reduce it to the one  we are considering.

The smallness of the group $G$ allows four antiunitary operators that commute with it: 1) $T_1 = i\beta_1 \beta_2 K$, $T_1 ^2 =-1$, 2)  $T_2 = K$, $T_2 ^2 =1$, 3) $T_3 = i\alpha_1 \alpha_2 K $, $T_3 ^2 = -1$, and 4) $T_4=\alpha_3 K$, $T_4^2 =1$ (Note that $\alpha_3 = \alpha_1 \alpha_2 \beta_1 \beta_2$). $T_3$ and $T_4$ happen to commute with the graphene Hamiltonian $H_1$, and since this Hamiltonian actually describes spinless lattice fermions, $T_4$ with a positive square represents the physical time reversal symmetry.\cite{herbut-clifford} $T_1$ and $T_2$ commute with $H_2$, and similarly $T_2$ is the physical time reversal.

The irreducible tensors in the $4 \times 4$ space under the symmetry are now: 1) $1_4$ (scalar-scalar), 2) $i\alpha_1 \alpha_2$ (scalar-scalar), 3) $i\beta_1 \beta_2$ (scalar-scalar), 4) $\alpha_3$ (scalar-scalar), 5) $(\alpha_1, \alpha_2)$ (vector-scalar), 6) $(\beta_1, \beta_2)$ (scalar-vector), 7) $(i\alpha_3 \alpha_1, i\alpha_3 \alpha_2)$  (vector-scalar), 8) $(i\alpha_3 \beta_1, i\alpha_3 \beta_2)$ (scalar-vector), and 9) $i \alpha_i \beta_j$, $i=1,2$, $j=1,2$ (vector-vector). We may now choose any of the four identified $T$-operators, and taking $T_2$ seems like the simplest choice: the tensors are even, odd, odd, even, even, odd, odd, even, even. Since the chosen $T_2 ^2 =1$ and $\Psi$ is Grassmann the number of independent terms matches the number of odd tensors, which is four. Taking $T_4$  also leads to four, albeit different, odd tensors, whereas taking either $T_1$ or $T_3$ yields four even tensors, and thus to the same conclusion. This agrees with explicit computation of the $9\times9$ Fierz matrix.\cite{vafek, rahul} The mixed terms are now forbidden by requiring both the time reversal and the parity to be extra symmetries.

  It is easy to see that adding the spin-1/2 degree of freedom
  to the problem and considering two four-component Grassmann fields $\Psi_\omega$ with $\omega =1,2$
and the group to be $G= SU(2) \times U(1) \times U(1)$ inevitably leads to nine independent couplings. Take, for example, 
the antiunitary Casimir operator to be
$T= \sigma_2 \otimes T_2 $, with $T^2=-1$. If the $4 \times 4$ operator $O$  was even under $T_2$, operator $1\otimes O$ will be even under $T$; if $O$ was odd under $T_2$, $\sigma_i \otimes O$ will also be even. The number of tensors even under $T$ is therefore the number of tensors even under $T_2$ plus the number of tensors odd under $T_2$, that is the total number of tensors before the spin doubling, which is nine. This again agrees with the explicit computation of the $18\times 18$ Fierz matrix.\cite{vafek}

\section{Conclusion}

 In conclusion, the hidden role of the antiunitary operators $T$ that commute with the symmetry group $G$  in a given representation in determining the independent local quartic terms in the field theory is revealed. When they exist, it suffices to simply count the T-even or T-odd tensors in the representation, to tell the number and often the identity of the independent terms. The crucial step in proving the above statement is the identification of type-II, or particle-particle, G-invariant quartic terms, which vanish identically. The theorem reproduces many results of explicit computations in the literature, and hopefully will aid and guide similar efforts in future.

\section{Acknowledgement}

This work was supported by the NSERC of Canada.

\end{document}